\documentclass[nonacm, sigconf]{acmart}

\usepackage{natbib}
\usepackage{multirow}
\usepackage{graphicx}
\usepackage{markdown}
\usepackage{subcaption}
\usepackage{multicol}
\usepackage{ulem}
\usepackage{listings}
\usepackage{varwidth}

\usepackage{todonotes}

\begin{document}

\title{Fairness for niche users and providers: \\algorithmic choice and profile portability}

\author{Elizabeth McKinnie}
\email{elizabeth.mckinnie@colorado.edu}
\orcid{0009-0002-8721-5700}
\affiliation{%
  \institution{Department of Information Science, University of Colorado, Boulder}
  \city{Boulder}
  \state{Colorado}
  \country{USA}
  \postcode{80309}
}

\author{Anas Buhayh}
\email{anas.buhayh@colorado.edu}
\orcid{0009-0009-7987-7967}
\affiliation{%
  \institution{Department of Information Science, University of Colorado, Boulder}
  \city{Boulder}
  \state{Colorado}
  \country{USA}
  \postcode{80309}
}

\author{Clement Canel}
\email{clement.canel@colorado.edu}
\orcid{}
\affiliation{%
  \institution{Department of Computer Science, University of Colorado, Boulder}
  \city{Boulder}
  \state{Colorado}
  \country{USA}
  \postcode{80309}
}

\author{Robin Burke}
\email{robin.burke@colorado.edu}
\orcid{0000-0001-5766-6434}
\affiliation{%
  \institution{Department of Information Science, University of Colorado, Boulder}
  \city{Boulder}
  \state{Colorado}
  \country{USA}
  \postcode{80309}
}

\begin{abstract}
Ensuring fair outcomes for multiple stakeholders in recommender systems has been studied mostly in terms of algorithmic interventions: building new models with better fairness properties, or using reranking to improve outcomes from an existing algorithm. What has rarely been studied is structural changes in the recommendation ecosystem itself. Our work explores the fairness impact of algorithmic pluralism, the idea that the recommendation algorithm is decoupled from the platform through which users access content, enabling user choice in algorithms. Prior work using a simulation approach has shown that niche consumers and (especially) niche providers benefit from algorithmic choice. In this paper, we use simulation to explore the question of profile portability, to understand how different policies regarding the handling of user profiles interact with fairness outcomes for consumers and providers.
\end{abstract}

\maketitle

\section{Introduction}
There is substantial literature on fairness-aware recommender systems, surveyed in \cite{ekstrand2022fairness}. This work focuses on two types of interventions: model-based (incorporating fairness objectives into the recommendation model) and re-ranking (applying post-hoc adjustments to recommendation lists). We are interested in exploring the fairness properties of a solution formulated at the ecosystem level, a market-based alternative proposed under various names: \textit{friendly neighborhood algorithm store}~\cite{rajendra2023three}, \textit{middleware}~\cite{fukuyama2021save,hogg2024shapingfuturesocialmedia}, or \textit{algorithmic pluralism}~\cite{verhulst2023steering}. The concept is that users should be able to choose among recommendation algorithms from different operators. Just as with other market niches, we might expect algorithm designers to specialize in serving particular audiences, and fairness — both consumer-side and provider-side — may be enhanced over solutions that rely on a single recommender system to be all things to all people. 

Using a simulation-based approach, \citet{buhayh2025decoupled} and \citet{buhayh2025simulating} explored the dynamics of a recommender system marketplace with two algorithm providers: one for a generic audience and one focused on a particular content niche. These studies found that users with distinct tastes outside of the mainstream (Niche consumers) and those who produced content catering to these tastes (Niche providers) were able to achieve better outcomes with minimal loss to the utility experienced by other consumers and providers. 

In this work, we use the SMORES model introduced in \cite{buhayh2025decoupled} to consider the impact of regulatory choices on the benefits achieved through an algorithmic market. In particular, we study the question of profile portability: how is user data shared when there are multiple algorithm operators? Since these operators compete with each other, they may not wish to hand over data collected from a particular user to a rival algorithm. This creates a significant degree of `lock-in': users may be reluctant to move if their data cannot be transferred with them. Therefore, one regulatory option would be to require profiles to be portable and to follow the user. We term this aspect of profile management \textit{exclusivity}: is user profile data tied to the recommender that gathered it?

The accumulation of data is crucial for the operation of large-scale machine learning platforms, such as recommender systems. Algorithm operators will want to retain as much data as they can, even if users move on. However, one could imagine a regulatory regime in which operators are required to delete a user's data if that user moves to a different system. This second aspect we term \textit{permanence}: whether user data is deleted when a user switches algorithms. 

\begin{figure*}
    \centering
    \includegraphics[width=0.75\textwidth]{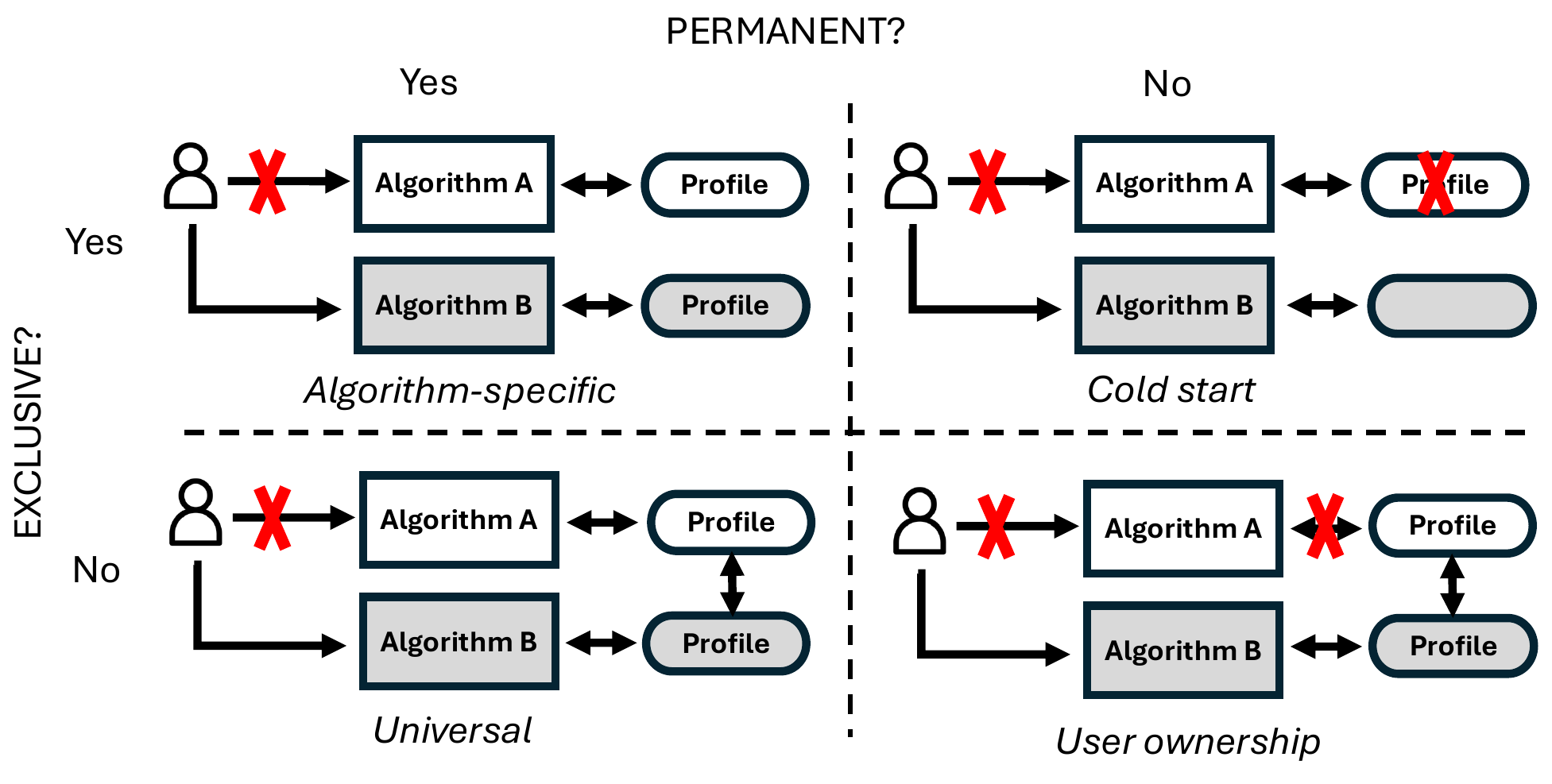}
    \caption{Different profile portability options.\protect\footnotemark}
    \label{fig:portability-options}
\end{figure*}

The interaction between exclusivity and permanence gives us four different conditions as depicted in Figure~\ref{fig:portability-options}.

\begin{itemize}
    \item \textbf{Algorithm-Specific} (Exclusive/Permanent, upper left): This condition corresponds to what we might expect in a competitive algorithmic market. Users have no control over their profiles, so algorithms are free to retain them when users switch. Algorithms are competitive, so they also do not allow data accumulated on their platform to be transferred to other platforms.
    \item \textbf{Cold Start} (Exclusive/Non-permanent, upper right): This is the most data-restrictive condition. Users cannot transfer their data from one algorithm to another, and when they switch away from a recommender, their profile is permanently deleted. Every user is, therefore, a cold-start user when switching algorithms.
    \item \textbf{User Ownership} (Non-exclusive/Non-permanent, lower right): In this condition, users' profiles move with them between algorithms, and so each profile is a mix of data gathered by all the algorithms with which the user has interacted. Algorithms do not have access to profiles for non-subscribers. Users essentially own their profiles and transfer all of their data when changing algorithms. 
    \item \textbf{Universal Profile} (Non-exclusive/Permanent, lower left): As in User Ownership, users bring their data with them. The difference is that algorithms retain access to a user's profile even when that user is not currently a subscriber. This means a user's information remains as training data available to support recommendations to other users. 
\end{itemize}

\footnotetext{Icons courtesy of The Noun Project.}

These conditions give rise to research questions that we explore in this work. Following the example of \cite{buhayh2025decoupled}, we concentrate on the simplified case where there are only two recommenders: one serving the general population -- we call this the Generic recommender $R^G$ -- and a second Niche recommender system $R^N$ that has been created to meet the needs of a subset of users for whom the generic one does not serve well.

\begin{itemize}
    \item \textbf{RQ1}: How does the exclusivity of a profile impact outcomes for consumers transitioning from one recommender to another? We might expect that having a profile tied to a particular recommender would mean that consumers who switch will receive lower quality recommendations. 
    \item \textbf{RQ2}: How does the permanence of a profile impact consumers who are not changing recommenders? If profiles are not permanent, then these consumers (e.g., Algorithm A consumers in the User Ownership case) receive recommendations generated using a smaller and potentially less diverse dataset once Niche consumers have left. 
    \item \textbf{RQ3}: How do these profile options impact the utility of the recommender systems for item providers? For example, we might expect consumers to switch more in the Cold Start condition, as it is more difficult for a recommender to build an effective profile. This could benefit providers with broader portfolios since Niche consumers spend more time with the Generic recommender $R^G$. 
\end{itemize}

\section{Related Work}
Research on multistakeholder recommender systems categorizes stakeholders on these platforms into three main groups: consumers, providers, and the platform itself \cite{burke2017multisided}. As noted in the studies of multisided platforms \cite{evans2016matchmakers}, such systems are sustainable if they provide utility to all parties involved. 

Studies on the fairness of recommender systems have shown that some individuals and groups are treated unfairly on these platforms, leading to disparities in utility across stakeholders and among stakeholder groups \cite{ekstrand2022fairness}. Simulation studies suggest that such algorithmic biases create feedback incentivizing providers to adapt their content to align with what the recommender will recommend \cite{hron2022modeling, yao2023bad}. This theoretical finding concurs with user studies of providers who report prioritizing what they believe the algorithm favors over their true creative intentions \cite{Choi2023-uk}.

To connect providers of Niche items with appropriate consumers, \citet{buhayh2025decoupled} suggest utilizing algorithmic choice as proposed by \cite{fukuyama2021save,rajendra2023three,hogg2024shapingfuturesocialmedia}. This approach enables consumers to select a recommender system that maximizes their utility. This work (extended in \cite{buhayh2025simulating}) found that both consumers and providers of Niche items may achieve higher utility under algorithmic choice.

In this paper, we adapt the simulation environment proposed in \cite{buhayh2025decoupled,buhayh2025simulating} to study profile portability under algorithmic choice. However, there are key differences in the methodology used here. First, instead of Surprise, we use Lenskit LKPY \footnote{https://github.com/lenskit/lkpy} as our recommendation generation platform for its flexibility in specifying recommendation pipelines. Second, instead of clustering consumers and generating recommendations for groups, we generate recommendations individually for each user. Third, we calculate provider utility as a function of user clicks rather than exposure. We believe these adaptations help make the simulation more realistic. 

\citet{Abel2011-he} demonstrate that profile portability strategies significantly enhance personalization while also mitigating the severity of the cold-start problem. Researchers have further explored the value of profile portability \cite{wischenbart2012user, Bouraga2016-mj} and associated challenges, including privacy \cite{Heitmann2010-xg}. There are various technological barriers and economic incentives that may deter platforms from implementing these strategies. Here, we concentrate on the potential benefits for consumer and provider stakeholders.

\section{SMORES}
The SMORES (Simulation Model for Recommender EcoSystems) framework~\cite{buhayh2025decoupled} is a simulation environment for modeling multistakeholder recommender systems. SMORES enables the study of complex interactions among consumers, providers, and recommendation algorithms under controlled conditions, making it suitable for evaluating ecosystem-level design questions such as profile portability and algorithmic fairness. It is an open-source system written in Python and available from GitHub. \footnote{https://github.com/that-recsys-lab/smores}

SMORES represents a recommender ecosystem with three primary stakeholders: \textit{consumers} ($j \in J$), who receive and act on recommendations; \textit{providers} ($v \in V$), who produce items ($i \in I$); and \textit{recommenders} ($k \in K$), which generate recommendation slates. Consumers interact with a single recommender at a time, selecting items from slates based on their preferences, which are modeled as normalized genre-based preference vectors. Providers gain utility through consumers choosing their items from the recommendation slates, while recommenders collect interaction data to refine recommendations. The framework supports dynamic consumer behavior, allowing consumers to switch between recommenders based on their satisfaction/utility, and tracks utility outcomes over time. By simulating these interactions, SMORES facilitates the evaluation of fairness and utility for diverse stakeholder groups, particularly underserved Niche consumers and providers.

Although SMORES is a simulation environment, it works from real data. Models of each consumer's preferences are compiled from a training dataset, resulting in a genre preference vector for each user. These preference vectors are used to calculate how the simulated consumer will react when presented with recommendations. Provider catalogs are built from the actual associations between providers (e.g. movie studios) and the items available for recommendation. For more details about these data-oriented aspects of SMORES, see \cite{buhayh2025decoupled}.

SMORES operates in discrete time steps, organized into \textit{days} and \textit{cycles}. Each day, consumers query their chosen recommender for a recommendation slate and select an item (or none, if all recommended items are poor matches), which updates their profile and utility. A cycle comprises multiple days (in our experiments, 10 days), after which consumers reassess their recommender choice based on utility. Recommenders generate slates using collaborative filtering algorithms, informed by consumer interaction histories, and can be configured to specialize in specific content areas (e.g., Niche genres). The framework’s flexibility allows for experimentation with various switching policies, data retention strategies, and utility models, making it ideal for studying fairness under algorithmic pluralism.

For the purpose of these experiments, we model user decision-making using a simple threshold technique. The utility of a recommender to a consumer is updated each day based on the recommendation list that is delivered. Let $\mu_{j,k,t}$ be the utility for a single recommendation list delivered by recommender $k$ to consumer $j$ on day $t$, defined as the average similarity of items on that list to the consumer's genre preferences. Each consumer computes their current utility for a recommender, \( \bar{\mu}_{j,k,t} \), with recency bias \( \beta \) (set to 2.0 in our experiments), favoring recent interactions: 

\begin{equation}
    \bar{\mu}_{j,k,t} = (\bar{\mu}_{j,k,t-1} \times \beta + \mu_{j,k,t}) / (1 + \beta) 
\end{equation}

Consumers have a satisfaction threshold $\tau$ (0.2 in our experiments). If the utility for the current recommender $k$, $\bar{\mu}_{j,k}$, falls below $\tau$, then the consumer will consider switching. But it will only switch if either the other recommender $k'$ has not been tried yet or if $\bar{\mu}_{j,k'}$ is greater than $\bar{\mu}_{j,k}$.

The simulation process is captured in pseudocode below. Note that a consumer $j$ has an "attached" recommender, $j.k$. Let $\hat{K}$ be the set of active recommenders.
\begin{lstlisting}[language=Python, caption=Smores Simulation Process, mathescape=true, showtabs=false, showspaces=false, showstringspaces=false]
for each cycle:
  for k in $\hat{K}$:
    train(k)
  for each day:
    for each consumer j:
      recs = j.k.recommendations(j)
      $\bar{\mu}_{j,k,t}$ = update_utility(j, recs)
      i = select_item(j, recs)
      V.update_utility(i)
      if $\bar{\mu}_{j,k,t}$ < $\tau$:
        for k' in $\hat{K} \ne j.k$:
          if $\bar{\mu}_{j,k',t}$ >= $\bar{\mu}_{j,k,t}$ 
                or k' not tried:
            j.k = k'
      manage_profile(j.data)
\end{lstlisting}

The \texttt{manage\_profile} function implements the appropriate profile management step for the profile portability condition active in the experiment, copying and deleting user profile data as appropriate. 

\section{Methodology}

This study extends the SMORES framework to investigate how profile portability policies -- exclusivity and permanence -- impact fairness outcomes for Niche and Generic consumers and providers in a recommender ecosystem with algorithmic choice. 

\subsection{Experimental Setup}

All consumer interactions are simulated, but we base the simulation behavior on actual data. We experiment here with a version of the MovieLens 1M dataset~\cite{harper2015movielens}. It contains 1 million ratings from 6,040 users across 3,706 movies, and is augmented with TMDb data for provider (studio) details. Horror, preferred by 1.47\% of consumers, is defined as the Niche genre.

Consumers are \textit{Niche} ($J^N$) if their highest-rated genre is the Niche genre, otherwise \textit{Generic} ($J^G$). Providers are \textit{Niche} ($V^N$) if their output has a majority of items in the Niche genre (6.93\% of MovieLens providers for Horror), otherwise \textit{Generic} ($V^G$).

We simulate an ecosystem with two recommenders: a \textit{Generic} recommender $R^G$ serving all content genres and a \textit{Niche} recommender $R^N$ specializing in the less popular Horror genre. Both recommenders are implemented using the ImplicitMF algorithm from Lenskit \footnote{https://lkpy.lenskit.org/stable/}$^{,}$\footnote{We expect that recommenders for different audiences would emerge organically, rather than being predefined as they are here, and plan to study such emergence in future work.}. Because the recommender needs data to compute recommendation slates, we implement two levels of fallback recommendations. When there are not enough consumers to make any predictions, each recommender samples from a list of 100 popular items to make cold start predictions. When a new consumer joins the recommender and has no associated profile, the recommender draws from a non-personalized popularity-based recommender using the data from its current users. For efficiency in simulation, we train recommenders only once per cycle. 

All consumers begin connected to $R^G$, a typical single recommender scenario. To allow interaction histories to form and to avoid cold-start noise, switching is disabled for the first two cycles. After this warm-up period, the decision model takes effect and consumers may switch between recommenders based on the utility they experience. We also run the simulation with $R^G$ as the only recommender to provide a baseline (non-switching) condition.

Each cycle spans 10 days, with each consumer receiving a daily slate of 10 items, selecting an item (or none) based on preferences. At the end of the cycle, there is an evaluation step where each consumer decides whether to retain the current recommender or to switch. We think of a cycle as a subscription period between the user and the recommender system. The simulation runs for 10 cycles to capture long-term dynamics. We found this provided sufficient time for the simulation to stabilize and for consumers to build enough confidence in the expected utility from both recommenders. Longer simulation periods did not result in significant changes in user/recommendation affiliation.  

\subsection{Profile Portability Conditions}

To address RQ1 (exclusivity) and RQ2 (permanence), we simulate the four profile portability conditions by controlling the data available to each recommendation algorithm:
\begin{description}
    \item [Algorithm-specific:] When a user switches from one algorithm to another, the old algorithm retains the profile, and no data is transferred to the new algorithm.
    \item [Cold Start:] When a user switches, their profile is deleted from the old algorithm's database and not communicated to the new algorithm.
    \item [User Ownership:] On switching, the old algorithm deletes the user's profile, but before doing so, transmits the profile to the new algorithm.
    \item [Universal:] All algorithms draw from the same dataset. 
\end{description}

\subsection{Evaluation}

As outcomes of our experiments, we report stakeholder utilities. Consumer utility is defined as the average relevance of the recommendation slate to the user's genre preferences. While multiple methods exist for evaluating consumer utility, we argue that genre-based relevance offers a meaningful and interpretable proxy, particularly when modeling user satisfaction in media recommendation contexts. We average this list utility for each consumer over the course of a cycle and then average over consumer types when reporting the results. 

Provider utility is defined by the number of clicks on an affiliated item by a consumer, a common e-commerce metric. Due to the long tail of item popularity, many items and providers are never recommended to any consumer in our simulations. This makes an average over providers a bit less useful as a metric, because the set of recommended providers may differ from scenario to scenario. Instead, we report on the total provider utility over all cycles, broken down by provider type. In future work, we plan to examine the distribution of provider utility in greater detail. 

For consumers, we focus on the utility achieved in the last simulation cycle when the distribution of consumers has (more or less) achieved a steady state. Unlike consumers, whose utility of the moment is the primary concern, providers are interested in long-term benefits, so we report cumulative clicks by provider type across the entire experiment time span. The trajectories of consumers across the simulation time span are of interest but are omitted here due to space constraints.

\section{Results}
The results of the study relative to consumers are shown in Table~\ref{tab:utilities_consumer_ml} and Figure~\ref{fig:utilities_consumer_ml}. Concentrating on Generic consumers, we see that there is not much difference across the various portability scenarios, with all scenarios being only slightly better than the non-switching Baseline. In the Figure, we observe that these differences do not reach significance when the distribution of values is taken into account. 

The Niche consumers, however, show a very different picture. Although all of the conditions are similar to each other, with an average utility somewhat higher than that of the Generic consumers, this average is much higher (more than 2x) above the Baseline. Corroborating the findings in \cite{buhayh2025simulating}, Niche consumers are better served when they can switch to a recommender tailored to their needs. 

\begin{table*}[th]
  \begin{varwidth}[b]{0.4\linewidth}
    \centering
\begin{tabular}{llr}
\toprule
Consumer Type & Scenario & Average Utility \\
\midrule
Generic & Baseline & 0.413 \\
 & Algorithm-Specific & 0.421 \\
 & Cold Start & 0.423 \\
 & User Ownership & 0.417 \\
 & Universal & 0.422 \\

\midrule
Niche & Baseline & 0.246 \\
 & Algorithm-Specific & 0.525 \\
 & Cold Start & 0.508 \\
 & User Ownership & 0.501 \\
 & Universal & 0.511 \\
\bottomrule
\end{tabular}
\caption{Average consumer utility (last cycle) across experimental conditions by consumer type.}
\label{tab:utilities_consumer_ml}
\end{varwidth}%
\hfill
\begin{minipage}[b]{0.5\linewidth}
    \centering
    \includegraphics[width=\linewidth, height=0.8\linewidth]{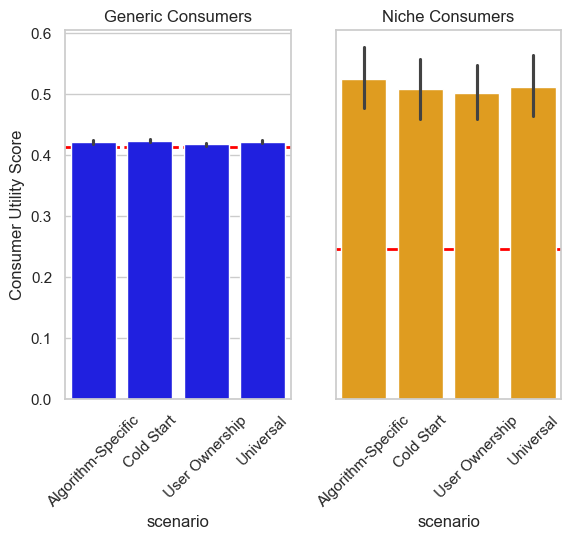}
    \captionof{figure}{Consumer utility (last cycle) across conditions by consumer type. Red dashed line shows consumer baseline for that consumer type.}
    \label{fig:utilities_consumer_ml}
\end{minipage}
\end{table*}

Interestingly, the exclusive conditions of Cold Start and Algorithm-Specific are not detrimental to Niche consumers. We note that the utility calculations for these consumers will tend to reward results in their specific genres. We'd expect that when the Niche recommender has less information about consumers' experiences with $R^G$, it is more likely to adhere to a narrow definition of the genre. Die-hard consumers in $R^N$ would then get more benefit from these focused recommendations. In the Universal and User Ownership cases, the Niche recommender would incorporate profile information for the (large majority) of Generic consumers whose tastes are not that useful in predicting for the Niche consumers. However, while User Ownership does the worst, the second-worst is Cold Start, and then Universal. The average utility values are very close together though (within 0.003 between Universal and Cold Start); to fully reject our hypothesis, we would need to conduct multiple experiments and calculate statistical significance of the findings.

Table~\ref{tab:utilities_provider_ml} and Figure~\ref{fig:utilities_provider_ml} show the results for providers. Recall that we are examining cumulative utility (clicks) across the entire experiment. We see that the switching conditions entail very little loss for the Generic providers (and in all cases except Algorithm-Specific, lead to gains in utility). That is different than the results found in \cite{buhayh2025decoupled,buhayh2025simulating}, which showed substantial loss for these providers. One factor could be our our utility calculation. In prior work, utility for providers was calculated in terms of exposure: any item shown to a consumer was considered valuable. However, this metric may overestimate the impact on Generic providers. In the Baseline condition, Niche consumers have nowhere to go and they will, of course, encounter the Generic content, which counts towards those providers' utility. The Generic providers lose out on their captive audience when switching is enabled. However, Niche consumers are mostly not interested and therefore do not select these items anyway. When we consider utility in terms of clicks, a consumer must be truly interested in the content for the provider to gain utility.

Interestingly, the Cold Start condition performs the best, and above Baseline, even though consumers who switch away from the recommender and then return have to rebuild their profiles from scratch and lose the benefits of personalization for some cycles. If niche consumers are staying with the niche recommender (which is confirmed by the data and discussed at the end of this section), the benefits of the recommender tuned to generic content as opposed to niche content could outweigh the loss of personalization. The Cold Start condition means that only the data gained in that recommender is used, and so if the Generic providers are already doing well there, they will continue to do well.

Generic providers in the Algorithm-Specific condition did perform worse, and below the Baseline value. Given that Cold Start did well, and the difference between the two conditions being that Algorithm-Specific retains data when consumers switch, this implies that the retained data did more harm than a clean slate did; perhaps Niche consumers were happy with some generic content, which then skewed future recommendations for Generic consumers.

For Niche providers, the benefits are very obvious. They derive very little utility from the Generic recommender, so the switching conditions are all in their favor. The one condition where Niche providers had the lowest average utility is Algorithm-Specific, similar to the case with Generic providers. The best performing condition is Universal, followed by User Ownership. These conditions allow for positive interactions that consumers had with niche items in a previous recommender to influence a different recommender that they use.

\begin{table*}[th]
  \begin{varwidth}[b]{0.4\linewidth}
    \centering
\centering
\begin{tabular}{llr}
\toprule
Provider Type & Scenario & Cumulative Utility \\
\midrule
Generic & Baseline & 13712.000 \\
 & Algorithm-Specific & 12719.000 \\
 & Cold Start & 15320.000 \\
 & User Ownership & 14193.000 \\
 & Universal & 14994.000 \\
\midrule
Niche & Baseline & 68.000 \\
 & Algorithm-Specific & 286.000 \\
 & Cold Start & 298.000 \\
 & User Ownership & 550.000 \\
 & Universal & 660.000 \\
\bottomrule
\end{tabular}
\caption{Cumulative Provider utility (all cycles) across experimental conditions by provider type.}
\label{tab:utilities_provider_ml}
\end{varwidth}%
\hfill
\begin{minipage}[b]{0.5\linewidth}
    \centering
    \includegraphics[width=\linewidth, height=0.8\linewidth]{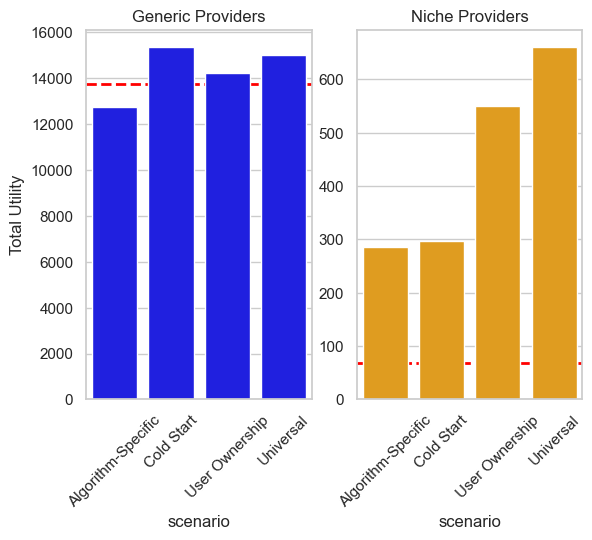}
    \captionof{figure}{Cumulative provider utility (all cycles) \\across conditions by provider type. Red dashed line shows \\results for the non-switching Baseline.}
    \label{fig:utilities_provider_ml}
\end{minipage}
\end{table*}

Figure~\ref{fig:switching} provides additional hints about the different portability conditions. It shows the number of times consumers switch to each of the recommenders, broken down by consumer type. For Generic consumers, more consumers switched to the Niche recommender than to the Generic recommender, across all profile conditions. The "To Niche" condition reflects dissatisfaction with the Generic recommender. We see it is highest in the Cold Start condition, which follows from the fact that the Cold Start condition has the least knowledge (after consumers have switched back). The Algorithm-Specific condition has the lowest amount of switching to the Niche recommender, which means that Generic consumers were more satisfied with their recommendations from the Generic recommender. The "To Generic" condition reflects the capabilities of the Niche recommender. The pattern differs from the Generic recommender in that Algorithm-Specific has the highest number of switching events, but this reinforces that the Generic recommender is better for Generic consumers, as Generic consumers become dissatisfied with the Niche recommender and switch back. User Ownership has the lowest number of switch actions, suggesting that consumers who bring their data with them to the Niche recommender then find the recommender to provide acceptable recommendations.

The figure shows that Niche consumers only ever switch to the Niche recommender; they never switch from the Niche recommender back to the Generic recommender, meaning they are satisfied with their recommendations from the Niche recommender. The number of consumers switching is about the same for each profile condition, showing that the condition has limited impact on switch behavior. This reinforces what we discovered earlier, that Niche consumers have higher utility when they have the choice to use a recommendation algorithm that's better tailored for them.

\section{Conclusion}
In this paper, we address the question of multistakeholder fairness in recommender systems through the lens of algorithmic choice. We simulate consumers' interactions with a recommender ecosystem in which consumers can choose among different recommendation algorithms, and study the impact of different profile portability policies. Our simulation simplifies the recommender ecosystem by predefining a set of consumers with particular non-mainstream tastes and assuming the existence of a recommender system tailored for them.

We note that our results are somewhat preliminary and need to be explored over a range of datasets and simulation parameters. Nonetheless, we can draw certain conclusions relative to our research questions. RQ1 asks whether profile portability impacts the quality of recommendations that consumers receive. At least in these studies, the answer is ``not much''. The biggest difference is between the availability of switching and its absence, and the (positive) impact is felt largely by the Niche consumers and providers. The Algorithm-Specific condition does seem linked to lower recommendation utility, but this effect is limited to providers.

\begin{figure}[bth]
    \centering
    \includegraphics[width=\linewidth]{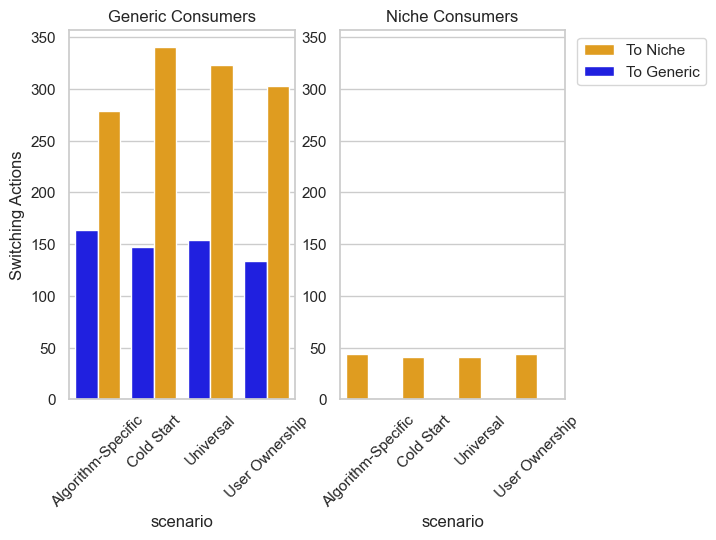}
    \caption{Switching behavior (all cycles) across portability conditions, by consumer type.}
    \label{fig:switching}
\end{figure}

RQ2 asks about the impact on consumers who are not switching recommenders. We know from Figure~\ref{fig:switching} that switching is relatively rare, so the average utilities measured for the consumers reflect this group. We see in most cases that the utility is higher. More mainstream consumers are not affected negatively if switching is enabled.   

RQ3 asks if profile portability impacts the utility of providers. Although Niche providers performed better in all switching conditions than Baseline, as we expected, it was true that they performed worse in low-information conditions such as Cold Start and Algorithm-Specific compared to high-information conditions. Generic providers benefited from most of the profile conditions, except for the Algorithm-Specific condition.

Interestingly, we find that the costs and benefits of different profile portability policies accrue to different groups of stakeholders in different ways. The Universal condition results in much higher utility for Niche providers than the Algorithm-Specific condition, but to Niche consumers, their utility is about the same. The Algorithm-Specific condition is also worse for Generic providers, but doesn't much impact the utility of Generic consumers. In considering the creation of such policies, regulators would need to balance stakeholder impacts appropriately. 

Further work is needed to study the effects suggested by these experiments. We plan to extend the work to different data sets and different choice strategies and mechanisms. The SMORES simulation is sufficiently modular to allow a wide variety of such explorations. We would also like to explore less constrained versions of the multi-algorithm recommender ecosystem, including those that would allow for the real-time emergence of recommenders in response to consumer needs. 

Finally, we note that despite the fairness benefits shown here, there are serious questions about the practical impacts of algorithmic choice for recommender systems. In an extended monograph on middleware in social media, \citet{hogg2024shapingfuturesocialmedia} outline some of these challenges. In particular, one could easily imagine a scenario in which algorithmic choice significantly reduces provider fairness through the emergence of highly exclusionary recommenders designed as echo chambers. One could imagine recommenders designed to surface harmful content or otherwise evade public safety guidelines that social media companies try to uphold. They proposed a number of solutions, including something in the nature of an ``app store'' where a base level of quality would be assured before an algorithm could become an option that users could choose. Clearly, more research in this area is needed before such systems can be designed and implemented.

\bibliographystyle{ACM-Reference-Format}
\bibliography{references}

\end{document}